\documentclass[superscriptaddress,twocolumn,secnumarabic,
amssymb,amsmath,nobibnotes,aps,prd,showkeys,showpacs,nofootinbib]{revtex4}%
\usepackage{graphicx}
\usepackage{epsf}
\usepackage{bm}
\usepackage{amsmath}
\usepackage{amsfonts}
\usepackage{amssymb}%
\providecommand{\U}[1]{\protect\rule{.1in}{.1in}}
\newcommand{\be}{\begin{equation}}
\newcommand{\ee}{\end{equation}}

\newcommand{\mincir}{\raise
-3.truept\hbox{\rlap{\hbox{$\sim$}}\raise4.truept\hbox{$<$}\ }}
\newcommand{\magcir}{\raise
-3.truept\hbox{\rlap{\hbox{$\sim$}}\raise4.truept\hbox{$>$}\ }}

\begin{document}
\title{Using the Noether symmetry approach to probe the nature of dark energy}
\author{Spyros Basilakos}
\affiliation{Academy of Athens, Research Center for Astronomy and Applied Mathematics,
Soranou Efesiou 4, 11527, Athens, Greece}
\author{Michael Tsamparlis}
\affiliation{Faculty of Physics, Department of Astrophysics - Astronomy - Mechanics
University of Athens, Panepistemiopolis, Athens 157 83, Greece}
\author{Andronikos Paliathanasis}
\affiliation{Faculty of Physics, Department of Astrophysics - Astronomy - Mechanics
University of Athens, Panepistemiopolis, Athens 157 83, Greece}

\begin{abstract}
We propose to use a model-independent criterion based on first integrals of
motion, due to Noether symmetries of the equations of motion, in order to
classify the dark energy models in the context of scalar field (quintessence
or phantom) FLRW cosmologies. In general, the Noether symmetries play an
important role in physics because they can be used to simplify a given system
of differential equations as well as to determine the integrability of the
system. The Noether symmetries are computed for nine distinct accelerating
cosmological scenarios that contain a homogeneous scalar field associated with
different types of potentials. We verify that all the scalar field potentials,
presented here, admit the trivial first integral namely energy conservation,
as they should. We also find that the exponential potential inspired from
scalar field cosmology, as well as some types of hyperbolic potentials,
include extra Noether symmetries. This feature suggests that these potentials
should be preferred along the hierarchy of scalar field potentials. Finally,
using the latter potentials, in the framework of either quintessence or
phantom scalar field cosmologies that contain also a non-relativistic matter
(dark matter) component, we find that the main cosmological functions, such as
the scale factor of the universe, the scalar field, the Hubble expansion rate
and the metric of the FRLW space-time, are computed analytically.
Interestingly, under specific circumstances the predictions of the exponential
and hyperbolic scalar field models are equivalent to those of the $\Lambda$CDM
model, as far as the global dynamics and the evolution of the scalar field are
concerned. The present analysis suggests that our technique appears to be very
competitive to other independent tests used to probe the functional form of a
given potential and thus the associated nature of dark energy.

\end{abstract}

\maketitle

\hyphenation{tho-rou-ghly in-te-gra-ting e-vol-ving con-si-de-ring
ta-king me-tho-do-lo-gy fi-gu-re}

\section{Introduction}

The analysis of the available high quality cosmological data (supernovae type
Ia, CMB, galaxy clustering, power spectrum etc) have converged during the last
decade towards a cosmic expansion history that involves a spatially flat
geometry and a recently initiated accelerated expansion of the universe (see
\cite{Teg04,Spergel07,essence,Kowal08,Hic09,komatsu08} and references
therein). From a theoretical point of view, an easy way to explain this
expansion is to consider an additional energy component, usually called dark
energy (DE) with negative pressure, that dominates the universe at late times.
The simplest DE candidate corresponds to the cosmological constant (see
\cite{Weinberg89,Peebles03,Pad03} for reviews). Indeed the so called spatially
flat concordance $\Lambda$CDM model, which includes cold dark matter (DM) and
a cosmological constant, ($\Lambda$) fits accurately the current observational
data and thus it is an excellent candidate to be the model which describes the
observed universe.

However, the concordance model suffers, among other \cite{Peri08}, from two
fundamental problems: (a) \textit{The ``old'' cosmological constant problem}
(or \textit{fine tuning problem}) i.e., the fact that the observed value of
the vacuum energy density ($\rho_{\Lambda}=\Lambda c^{2}/8\pi G\simeq
10^{-47}\,GeV^{4}$) is many orders of magnitude below the value found using
quantum field theory \cite{Weinberg89}, and (b) \textit{the coincidence
problem}\,\cite{coincidence} i.e., the fact that the matter energy density and
the vacuum energy density are of the same order (just prior to the present
epoch), despite the fact that the former is a rapidly decreasing function of
time while the latter is stationary.

Attempts to solve the coincidence problem have been presented in the
literature (see \cite{Peebles03,Pad03,Egan08} and references therein), in
which an easy way to overpass the coincidence problem is to replace the
constant vacuum energy with a DE that evolves with time. Nowadays, the physics
of DE is considered one of the most fundamental and challenging problems on
the interface uniting Astronomy, Cosmology and Particle Physics. In the last
decade there have been theoretical debates among the cosmologists regarding
the nature of this exotic component. Many candidates have been proposed in the
literature, such as a cosmological constant $\Lambda$ (vacuum), time-varying
$\Lambda(t)$ cosmologies, quintessence, $k-$essence, quartessence, vector
fields, phantom, gravitational matter creation, tachyons, modifications of
gravity, Chaplygin gas and the list goes on (see
\cite{Ratra88,Oze87,Weinberg89,Lambdat,Bas09c,Wetterich:1994bg,
Caldwell98,Brax:1999gp,KAM,fein02,Caldwell,Bento03,chime04,Linder2004,LSS08,
Brookfield:2005td,Grande06,Boehmer:2007qa} and references therein).

In the original scalar field models\,\cite{Dolgov82} and later in the
quintessence context, one can ad-hoc introduce an adjusting or tracker scalar
field $\phi$\,\cite{Caldwell98} (different from the usual SM Higgs field),
rolling down the potential energy $V(\phi)$, which could resemble the DE
\cite{Peebles03,Pad03,Jassal,SR,Xin,SVJ}. However, it was realized that the
idea of a scalar field rolling down some suitable potential does not really
solve the problem because $\phi$ has to be some high energy field of a Grand
Unified Theory (GUT), and this leads to an unnaturally small value of its
mass, which is beyond all conceivable standards in Particle Physics. As an
example, utilizing the simplest form for the potential of the scalar field,
$V(\phi)=m_{\phi}^{2}\,\phi{^{2}}/2$, the present value of the associated
vacuum energy density is $\rho_{\Lambda}=\langle V(\phi) \rangle\sim
10^{-11}\,eV^{4}$, so for $\langle\phi\rangle$ of order of a typical GUT scale
near the Planck mass, $M_{P}\sim10^{19}$ GeV, the corresponding mass of $\phi$
is expected in the ballpark of $m_{\phi}\sim\,H_{0}\sim10^{-33}\,eV$.

Notice that the presence of such a tiny mass scale in scalar field models of
DE is generally expected also on the basis of structure formation
arguments\,\cite{Mota04,Nunes06,Basi09}; namely from the fact that the DE
perturbations seem to play an insignificant role in structure formation for
scales well below the sound horizon. The main reason for this homogeneity of
the DE is the flatness of the potential, which is necessary to produce a
cosmic acceleration. Being the mass associated to the scalar field fluctuation
proportional to the second derivative of the potential itself, it follows that
$m_{\phi}$ will be very small and one expects that the magnitude of DE
fluctuations induced by $\phi$ should be appreciable only on length scales of
the order of the horizon. Thus, equating the spatial scale of these
fluctuations to the Compton wavelength of $\phi$ (hence to the inverse of its
mass) it follows once more that $m_{\phi}\lesssim\,H_{0}\sim10^{-33}\,eV$.

Despite the above difficulties there is a class of viable models of
quintessence based on supersymetry, supergravity and string-theory which can
protect, under of specific potentials, the light mass of quintessence (for a
review see \cite{Ame10} and references therein). In spite of that, this class
of DE models have been widely used in the literature due to their simplicity.
Notice that DE models with a canonical kinetic term have a dark energy EoS
parameter $-1\leq w_{\phi}<-1/3$. Models with ($w_{\phi}<-1$), sometimes
called phantom DE \cite{phantom}, are endowed with a very exotic nature, like
a scalar field with negative kinetic energy. In any case, in order to
investigate the overall dynamics of the universe we need to define the
functional form of the potential energy. As we have already mentioned the
issue of the potential energy has a long history in scalar field cosmology and
indeed several parametrizations have been proposed (exponential, power law,
hyperbolic etc).

The aim of the present work is to investigate which of the available scalar
field potentials can accommodate basic geometrical symmetries (connected to
the space-time) namely Lie point and Noether. In fact the idea to use Noether
symmetries as a cosmological tool is not new. In particular, it has been
proposed that the existence of such symmetries are related with conserved
quantities and thus they can be used as a selection criterion in order to
discriminate the dark energy models, including those of $f(R)$ gravity (see
\cite{Cap96},\cite{Szy06},\cite{Cap07},\cite{Capa07},\cite{Cap08}%
,\cite{Cap09},\cite{Vakili08},\cite{Yi09}). From a mathematical point of view,
the Lie point/Noether symmetries play a vital role in physical problems
because they provide Noether (first) integrals, which can be used in order to
simplify a given system of differential equations and to determine the
integrability of the system. A fundamental approach to derive the Lie point
and Noether symmetries for a given dynamical problem moving in a Riemannian
space has been proposed recently by Tsamparlis \& Paliathanasis \cite{Tsam10}
(a similar analysis can be found in \cite{Leach},\cite{Blum}, \cite{Aminova
1994},\cite{Aminova 1995},\cite{Aminova 2000}, \cite{Feroze Mahomed
Qadir},\cite{Tsamparlis2010},\cite{Tsama10}).

The structure of the article is as follows. The basic theoretical elements of
the problem are presented in section 2, where we also introduce the basic FLRW
cosmological equations for various potentials of the scalar field. The
geometrical symmetries of the scalar fields and their connections to the
potential energy are discussed in section 3. In section 4 we provide for a
first time (to our knowledge) analytical solutions in the light of either
quintessence or phantom scalar field cosmologies that include non-relativistic
matter (dark matter). Finally, the main conclusions are summarized in section 5.

\section{Cosmology with a scalar field}

The scalar field contribution to the curvature of space-time can be absorbed
in Einstein's field equations as follows:
\begin{equation}
R_{\mu\nu}-\frac{1}{2}g_{\mu\nu}R=k\ \tilde{T}_{\mu\nu}\;\;\;\;\; k=8\pi G
\label{EE}%
\end{equation}
where $R_{\mu\nu}$ is the Ricci tensor and $\tilde{T}_{\mu\nu}$ is the total
energy momentum tensor $\tilde{T}_{\mu\nu}$ given by $\tilde{T}_{\mu\nu}\equiv
T_{\mu\nu}+T_{\mu\nu}(\phi)$. Here $T_{\mu\nu}(\phi)$ is the energy-momentum
tensor associated with the scalar field $\phi$, and $T_{\mu\nu}$ is the
ordinary energy-momentum tensor of matter and radiation. Modeling the
expanding universe as a perfect fluid that includes radiation, matter and DE
with $4-$velocity $U_{\mu}$, we have $\tilde{T}_{\mu\nu}=-P\,g_{\mu\nu}+
(\rho+P)U_{\mu}U_{\nu}$, where $\rho=\rho_{m}+\rho_{\phi}$ and $P=P_{m}%
+P_{\phi}$ are the total energy density and pressure of the cosmic fluid
respectively. Note that $\rho_{m}$ is the proper isotropic density of
matter-radiation, $\rho_{\phi}$ denotes the density of the scalar field and
$P_{m}$, $P_{\phi}$ are the corresponding pressures. In the context of a FLRW
metric with Cartesian coordinates
\begin{equation}
ds^{2}=-dt^{2}+a^{2}(t)\frac{1}{(1+\frac{K_{3}}{4}\mathbf{x}^{2})^{2}}%
(dx^{2}+dy^{2}+dz^{2}) \label{SF.1}%
\end{equation}
the Einstein's field equations (\ref{EE}), for comoving observers ($U^{\mu
}=\delta_{0}^{\mu}$), provide
\begin{align}
R_{00}  &  =-3\frac{\ddot{a}}{a}\label{SF.2}\\
R_{\mu\nu}  &  =\left[  \frac{\ddot{a}}{a}+2\frac{\dot{a}^{2}+K_{3} }{a^{2}%
}\right]  g_{\mu\nu} \label{SF.3a}%
\end{align}
where the over-dot denotes derivative with respect to the cosmic time $t$,
$a(t)$ is the scale factor of the universe, and $K_{3}=0,\pm1$ is the spatial
curvature parameter. Also, the contraction of the Ricci tensor provides the
Ricci scalar
\begin{equation}
R=g^{\mu\nu}R_{\mu\nu}= 6\left[  \frac{\ddot{a}}{a}+\frac{\dot{a}^{2}+K_{3}%
}{a^{2}}\right]  . \label{SF.3b}%
\end{equation}
Finally, the gravitational field equations boil down to Friedmann's equation
\begin{equation}
H^{2}\equiv\left(  \frac{\dot{a}}{a}\right)  ^{2}=\frac{k}{3} \left(  \rho_{m}
+\rho_{\phi}\right)  -\frac{K_{3}}{a^{2}}, \label{frie1}%
\end{equation}
and
\begin{equation}
3H^{2}+2\dot{H}=-k(P_{m}+P_{\phi})-\frac{K_{3}}{a^{2}} \label{frie2}%
\end{equation}
where $H(t)\equiv\dot{a}/a$ is the Hubble function.
The Bianchi identity (which insure the covariance of the theory)
$\bigtriangledown^{\mu}\,{\tilde{T}}_{\mu\nu}=0$ amounts to the following
generalized local conservation law:%

\begin{equation}
\dot{\rho}_{m}+\dot{\rho_{\phi}}+ 3H(\rho_{m}+P_{m}+ \rho_{\phi}+P_{\phi
})=0\,. \label{frie3}%
\end{equation}
Note that the latter quantities obey the following relations:
\begin{equation}
(\rho_{m},P_{m})\equiv(-T_{0}^{0},T_{i}^{i}) \;\;\;\; (\rho_{\phi},P_{\phi
})\equiv(-T_{0}^{0}(\phi),T_{i}^{i}(\phi)) \;\;. \label{cons}%
\end{equation}
Combining eqs. (\ref{frie1}), (\ref{frie2}) and (\ref{frie3}) we obtain
\begin{equation}
\frac{\ddot{a}}{a}=-\frac{k}{6}\, [\rho_{m}+\rho_{\phi}+3\,(P_{m}+P_{\phi})].
\label{2FL}%
\end{equation}
Assuming negligible interaction between matter and scalar field, eq.
(\ref{frie3}) leads to the following independent differential equations
\begin{equation}
\dot{\rho}_{m}+ 3H(\rho_{m}+P_{m})=0\,, \label{frie4}%
\end{equation}
\begin{equation}
\dot{\rho}_{\phi}+ 3H(\rho_{\phi}+P_{\phi})=0\,. \label{frie5}%
\end{equation}
In this work, we will present the global dynamics of the universe in the
presence of a barotropic cosmic fluid whose the corresponding EoS parameters
are given by $w_{m}=P_{m}/\rho_{m}$ and $w_{\phi}=P_{\phi}/\rho_{\phi}$. In
what follows we assume a constant $w_{m}$ which implies that $\rho_{m}=\rho_{m
0} a^{-3(1+w_{m})}$ (cold $w_{m}=0$ and relativistic $w_{m}=1/3$ matter),
where $\rho_{m 0}$ is the matter density at the present time. Generically,
some high energy field theories suggest that the dark energy EoS parameter is
a function of cosmic time (see, for instance, \cite{Ellis05}) and thus
\begin{equation}
\rho_{\phi}(a)=\rho_{\phi0}\;{\exp}\left(  \int_{a}^{1} \frac{3[1+w_{\phi
}(\sigma)]}{\sigma} d\sigma\right)  \label{frie55}%
\end{equation}
where $\rho_{\phi0}$ is the DE density at the current epoch.

\subsection{The scalar field}

We consider a scalar field in the FRLW cosmology which is minimally coupled to
gravity, such that the field satisfies the Cosmological Principle that is,
$\phi$ inherits the symmetries of the metric. This means that the scalar field
depends only on the cosmic time $t$, and consequently $\phi_{,\nu}=\dot{\phi
}\delta_{\nu}^{0}$ where $\dot{\phi}=\frac{d\phi}{dt}$. A scalar field
$\phi(t)$ with a potential $V(\phi)$ is defined by the energy momentum tensor
of the form (for review see \cite{Ame10} and references therein)
\begin{equation}
T_{\mu\nu}(\phi)=-\frac{2}{\sqrt{-g}} \frac{\delta(\sqrt{-g}L_{\phi})}{\delta
g^{\mu\nu}} \label{tensor2}%
\end{equation}
where $L_{\phi}$ is the Lagrangian of the scalar field. Although in the
current analysis we study generically, as much as possible, the problem we
will focus on a scalar field with
\begin{equation}
L_{\phi}=-\frac{1}{2}\epsilon g^{\mu\nu}\phi_{,\mu}\phi_{,\nu}-V(\phi)
\label{SF.32}%
\end{equation}
or
\begin{equation}
L_{\phi}=\frac{1}{2}\epsilon\dot{\phi}^{2}-V(\phi) \label{Lag1}%
\end{equation}
where
\begin{equation}
\epsilon=\left\{
\begin{array}
[c]{cc}%
1 & \mbox{Quintessence}\\
-1 & \mbox{Phantom\;.}
\end{array}
\right.
\end{equation}
Therefore, using the second equality of eq.(\ref{cons}), eq.({\ref{tensor2})
and eq.({\ref{Lag1}) the energy density $\rho_{\phi}$ and the pressure
$P_{\phi}$ of the scalar field are given by
\begin{equation}
\rho_{\phi}\equiv-T_{0}^{0}(\phi)=\frac{1}{2}\epsilon\dot{\phi}^{2}+V(\phi)
\label{den1}%
\end{equation}
and
\begin{equation}
P_{\phi}\equiv T_{i}^{i}(\phi)=L_{\phi}= \frac{1}{2}\epsilon\dot{\phi}%
^{2}-V(\phi) \;\;. \label{pres1}%
\end{equation}
Inserting eq.(\ref{den1}) and eq.(\ref{pres1}) into eq.(\ref{frie5}) it is
routine to derive the Klein-Gordon equation which describes the time evolution
of the scalar field. This is
\begin{equation}
\ddot{\phi}+\frac{3}{a}\dot{a}\dot{\phi}+\epsilon V_{,\phi}=0 \label{klein1}%
\end{equation}
where $V_{,\phi}=dV/d\phi$. Obviously, if we use the current functional form
of $L_{\phi}$ then eq.(\ref{frie2}) takes the form:
\begin{equation}
\frac{\ddot{a}}{a}+\frac{1}{2}\left(  \frac{\dot{a}^{2}}{a^{2}}+ \frac{K_{3}%
}{a^{2}}\right)  +\frac{k}{2}\left(  P_{m}+\frac{1}{2}\epsilon\dot{\phi}%
^{2}-V(\phi)\right)  =0 \label{klein2}%
\end{equation}
}}

The corresponding dark energy EoS parameter (defined before) is
\begin{equation}
w_{\phi}=\frac{P_{\phi}}{\rho_{\phi}}=\frac{\epsilon(\dot{\phi}^{2}%
/2)-V(\phi)} {\epsilon(\dot{\phi}^{2}/2)+V(\phi)} \;\;. \label{pres11}%
\end{equation}
The quintessence ($\epsilon=1$) cosmological model accommodates a late time
cosmic acceleration in the case of $w_{\phi}<-1/3$ which implies that
$\dot{\phi}^{2}<V(\phi)$. On the other hand, if the kinetic term of the scalar
field is negligible with respect to the potential energy [$\frac{\dot{\phi
}^{2}}{2} \ll V(\phi)$] then the equation of state parameter is $w_{\phi
}\simeq-1$. In the case of a phantom DE ($\epsilon=-1$), due to the negative
kinetic term, one has $w_{\phi}<-1$ for $(\dot{\phi}^{2}/2)<V(\phi)$.

The unknown quantities of the problem are $a(t),$ $\phi(t)$ and $V(\phi)$ but
we have only two independent differential equations available namely eqs.
(\ref{klein1}) and (\ref{klein2}). Thus, in order to solve this system of
differential equations we need to assume a functional form of the scalar field
potential energy, $V(\phi)$. In the literature, due to the unknown nature of
the DE, there are many forms of potentials (for a review see \cite{Ame10})
which describe differently the physical features of the scalar field. Let us
now briefly present various potentials whose free parameters can be
constrained by using the current cosmological data.

\begin{itemize}
\item The power law potential \cite{Ratra88,Caldwell98}:
\begin{equation}
V(\phi)=\frac{M^{4+n}}{\phi^{n}} \label{pot1}%
\end{equation}

\item The exponential potential \cite{Sievers03}:
\begin{equation}
V(\phi)=V_{0}\;\mathrm{exp}(-\sqrt{k}\;\lambda\phi) \label{pot2}%
\end{equation}

\item The unified dark matter potential (hereafter UDM) \cite{Bertaca07}:
\begin{equation}
V(\phi)=V_{0}\;[1+\mathrm{cosh^{2}}(\lambda\phi)]\ \label{unpot2}%
\end{equation}

\item The Pseudo-Nambu Goldstone Boson potential \cite{Friem95}:
\begin{equation}
V(\phi)=\mu^{4}[1+\mathrm{cos}(\phi/f)] \label{pot3}%
\end{equation}

\item The exponential with inverse power \cite{stein99}:
\begin{equation}
V(\phi)=M[\mathrm{exp}(\gamma/\phi)-1] \label{pot4}%
\end{equation}

\item The supergravity motivated potential \cite{Brax:1999gp}:
\begin{equation}
V(\phi)=M \mathrm{exp}(\phi^{2})/\phi^{\gamma} \label{pot7}%
\end{equation}

\item The early dark energy potential \cite{Bare00}:
\begin{equation}
V(\phi)=V_{1}\;\mathrm{exp}(-\sqrt{k}\;\lambda_{1} \phi)+ V_{2}\;\mathrm{exp}%
(-\sqrt{k}\;\lambda_{2} \phi) \;. \label{pot5}%
\end{equation}
Interestingly the potential:
\[
V(\phi)=V_{0}[\mathrm{cosh}(\lambda\phi)-1]^{p}%
\]
provides predictions which are similar to those of the early DE model, as far
as the global dynamics of the universe is concerned \cite{Saa}.

\item The Albrecht-Skordis model \cite{Alb00}:
\begin{equation}
V(\phi)=M[A+(\phi-B)^{2}]\;\mathrm{exp}(-\gamma\phi) \label{pot6}%
\end{equation}

\item The Chaplygin gas from the ordinary scalar field viewpoint
\cite{Ame10}:
\begin{equation}
V(\phi)=\frac{\sqrt{A}}{2}\left(  \mathrm{cosh}\sqrt{3}k\phi+ \frac
{1}{\mathrm{cosh}\sqrt{3}k\phi} \right)  \;. \label{pot7}%
\end{equation}

\end{itemize}

Detailed analysis of these potentials exist in the literature, including their
confrontation with the data (see \cite{Ame10} for extensive reviews). It is
worth pointing out that for some special cases attempts to find analytical
solutions can be found in \cite{Turner83,mata85,stein99,santi00,sen02,
kehagias04,Russo04,Gorini05} (and references therein).

\section{Dark Energy versus space-time symmetries}

In the last decade, a large number of experiments have been proposed in order
to constrain DE and study its evolution. Naturally, in order to establish the
evolution of the DE equation of state a realistic form of $H(a)$ is required
while the included free parameters must be constrained through a combination
of independent DE probes (for example SNIa, BAOs, CMB etc). However, there is
always a range of parameters of the considered scalar field cosmological
models for which a good fit with the observational data is provided. This
implies that such DE models can not be distinguished observationally, since
they provide similarly evolving Hubble functions.

Practically, the goal here is to define a method (selection criterion) that
can distinguish the DE models on a more fundamental (eg. geometrical) level.
According to the theory of general relativity, the space-time symmetries
(Killing and homothetic vectors) via the Einstein's field equations (see
eq.\ref{EE}), are also symmetries of the energy momentum tensor (the matter
generates the gravitational field). Owing to the fact that the scalar field is
minimally coupled to gravity and it evolves in space-time one would expect
that the scalar field must inherit the symmetries of the space-time as gravity does.

It is interesting to mention that besides the geometric symmetries one has to
consider the dynamical symmetries, which are the symmetries of the field
equations. These latter symmetries are known as Lie symmetries. In case the
field equations are derived from a Lagrangian there is a special class of Lie
symmetries, the Noether symmetries, which have the characteristic that lead to
conserved currents or, equivalently, to first integrals of the equations of
motion. The Noether integrals are used to reduce the order of the field
equations or even to solve them. Therefore a sound requirement, which is
possible to be made in Lagrangian theories is that they admit extra Noether
symmetries. This assumption is model independent, because it is imposed after
the field equations have been derived, therefore it does not lead to conflict
whereas with the geometric symmetries while, at the same time, serves the
original purpose of a selection rule. Since the basic equations in the scalar
field cosmologies follow from a Lagrangian we can apply the above ideas by
looking for scalar field cosmologies which admit extra Noether symmetries.


In particular, let us consider a cosmological model which accommodates a late
time accelerated expansion and it contains a scalar field $\phi(t)$, described
by a potential $V(\phi)$. We pose the following question: \textit{For the
scalar field that lives into a 2-dimensional Riemannian space $\{a,\phi\}$ and
which is embedded in the space-time, how many (if any) of the previously
presented potentials (see section 2.1) can provide non trivial Noether
symmetries (or first integrals of motion})? As an example, if we find a
cosmological model (or a family of models) for which its scalar field produces
non-trivial number of first integrals of motion with respect to the other DE
cosmological models, then obviously this model contains an extra geometrical
feature. Thus, we can use this geometrical characteristic in order to classify
the explored DE cosmological model into a distinct category (see also
\cite{Cap96},\cite{Szy06},\cite{Cap07},\cite{Capa07},\cite{Cap08}%
,\cite{Cap09}). Below, we present the geometrical method used in order to
define the basic symmetries, namely Lie point and Noether which lead to first
integrals of motion.

\subsection{Lie and Noether symmetries}

We briefly present the main points of the method used to classify all two
dimensional Newtonian dynamical systems, which admit Lie point/Noether
symmetries\footnote{Note that the Noether symmetries are a sub-algebra of the
algebra defined by the Lie symmetries \cite{Tsam10}. A dot over a symbol in
eq.(\ref{L2P.1}) indicates derivation with respect to the parameter $s$ (the
affine parameter along the trajectory - cosmic time in our case)}. A first
important ingredient is the use of two theorems which relate the Lie point and
the Noether symmetries of a dynamical system moving in a Riemannian space with
the special projective group and the homothetic group generators of the space
respectively. These theorems have been given by some of us in \cite{Tsam10}.
In particular in a recent paper Tsamparlis \& Paliathanasis \cite{Tsam10} have
provided an alternative way to solve the system of Lie point/Noether symmetry
conditions (see their tables 1-15), for second order equations of the form:%
\begin{equation}
\ddot{x}^{i}+\Gamma_{jk}^{i}\dot{x}^{j}\dot{x}^{k}=F^{i}. \label{L2P.1}%
\end{equation}
Here $\Gamma_{jk}^{i}(x^{r})$ are general functions, along the solution curves
and $F^{i}(x^{j})$ is a $C^{\infty}$ vector field. Basically, equations
(\ref{L2P.1}) are the equations of motion of a dynamical system in a
Riemannian space in which the functions $\Gamma_{jk}^{i}(x^{r})$ are the
connection coefficients (Christofell symbols) of the metric $\hat{g}_{ij}$ of
the space (in our case $\{a,\phi\}$ see below). The key point, (see
\cite{Tsam10}), is to express the system of Lie point/Noether symmetry
conditions of eq.(\ref{L2P.1}) in terms of collineation (usually referred as
symmetries) conditions of the metric. If this is achieved, then the Lie
point/Noether symmetries of eq.(\ref{L2P.1}) will be related to the
collineations of the metric. Therefore the determination of the Lie
point/Noether symmetries of eq.(\ref{L2P.1}) will be transferred to the
geometric problem of determining the generators of a specific type of
collineations of the metric. Then it will be possible to use the plethora of
results of Differential Geometry on collineations to produce the solution of
the Lie point/Noether symmetry problem.

The natural question to ask is: \emph{How one will select the Lie
point/Noether symmetries of two different dynamical systems, which move in the
same Riemannian space?\ } The answer is simple. The left hand side of
eq.(\ref{L2P.1}) contains the metric and its derivatives and it is
\emph{common to all} dynamical systems moving in the same Riemannian space.
Therefore geometry enters in the left hand side of eq.(\ref{L2P.1}) only. Each
dynamical system is defined by the force field $F^{i},$ which enters into the
right hand side of eq.(\ref{L2P.1}) only. Therefore, there must exist
constraints, which involve the components of the Lie point/Noether symmetry
vectors and the force field, and which will have to be satisfied for a
collineation of the metric in order for it to be a Lie point/Noether symmetry
of the specific dynamical system. A similar approach can be found in
\cite{Aminova 1994},\cite{Aminova 1995}\cite{Aminova 2000},\cite{Feroze
Mahomed Qadir},\cite{Tsamparlis2010},\cite{Tsama10}.

In a subsequent work \cite{Tsam10} Tsamparlis \& Paliathanasis determined
(among others) all two dimensional potentials which admit at least one Lie
point and/or Noether symmetry. The results of the calculations have been
collected for convenience in tables 1-15. From these tables one can read
directly the aforementioned potentials and, furthermore, for each potential
the admitted Lie point and Noether symmetries together with the corresponding
Noether integrals. It is emphasized that no extra calculations are required.
In the following we shall make use of these results.

\subsection{Using the geometry of the space $\{a,\phi\}$ to constrain Dark
Energy}

In this section we apply the previous theory into the scalar field cosmology.
Interestingly, we can easily prove that the main field equations
(\ref{klein1}) and (\ref{klein2}), described in section 2, can be produced by
the following general Lagrangian:
\begin{equation}
L=-3a\dot{a}^{2}+ka^{3}L_{\phi}+ka^{3}P_{m}+3K_{3} a \label{SF.50}%
\end{equation}
in the space of the variables $\{a,\phi\}$. Also the action is
\begin{equation}
S=\int L \;d^{3}x\;dt \;. \label{action}%
\end{equation}
Therefore, utilizing eq.(\ref{Lag1}) and eq.(\ref{SF.50}) we can also obtain
the Hamiltonian of this system
\begin{equation}
E=\frac{1}{2}\left(  -6a\dot{a}^{2}+ka^{3}\epsilon\dot{\phi}^{2}\right)
+ka^{3}\left[  V(\phi)-P_{m}\right]  -3K_{3} a\;. \label{SF.60e}%
\end{equation}
In order to compute the Lie point/Noether symmetries of equations of motion
(\ref{klein1}) and (\ref{klein2}), we consider the Lagrangian as the sum of a
kinetic energy which defines the metric in the space of $\{a,\phi\}$ and an
external force field. In particular, this 2 dimensional metric takes the form
\begin{equation}
d\hat{s}^{2}=-6ada^{2}+ka^{3}\epsilon d\phi^{2} \label{SF.52}%
\end{equation}
which implies that
\begin{equation}
\Gamma_{aa}^{a}=\frac{1}{2a}~,\Gamma_{a\phi}^{\phi}=\frac{3}{2a},~\Gamma
_{\phi\phi}^{a}=\frac{k}{4}\epsilon a \;. \label{SF.53}%
\end{equation}
Obviously, in the case of phantom cosmology $\epsilon=-1$, eq.(\ref{SF.52})
points that that we have to replace $\phi$ with $i\phi$ for mathematical
convenience. Using now eq.(\ref{SF.53}) and inserting the variables
$x^{i}=a(t)$, $\phi(t)$ variables into eq.(\ref{L2P.1}) we find that
\begin{align*}
\ddot{a}+\frac{1}{2a}\dot{a}^{2}+ \frac{k}{4}a\epsilon\dot{\phi}^{2}  &
=F^{a}\\
\ddot{\phi}+\frac{3}{a}\dot{a}\dot{\phi}  &  =F^{\phi}.
\end{align*}
Comparing with the equations of motion (\ref{klein1}) and (\ref{klein2}), we
can define the external ''forces'' in terms of the scalar field potential
$V(\phi)$
\[
F^{a}=-\frac{1}{2a}K_{3}-\frac{ka}{2}\left[  P_{m}-V(\phi)\right]  ~,~F^{\phi
}=-\epsilon V_{,\phi}%
\]

On the other hand using the above $\Gamma_{jk}^{i}$ functions we find after
some simple algebra that the curvature of the $\{a,\phi\}$ space is $\hat
{R}=0$ implying flatness (all $2$ dimensional spaces are Einstein spaces hence
$\hat{R}=0$ implies the space is flat). Also, the signature of the metric
eq.(\ref{SF.53}) is $-1$, hence the space is the 2-d Minkowski space.
Therefore, according to theorems 1 and 2 of \cite{Tsam10} the Lie
point/Noether symmetries of the equations (\ref{klein1}) and (\ref{klein2})
follow from the special projective group of the 2-d Minkowski space. To find
explicitly these vectors and thus the corresponding Noether symmetries (or
first integrals), we have to bring the 2-d metric of eq.(\ref{SF.52}) into its
canonical form (i.e. $d\hat{s}^{2}=-dx^{2}+dy^{2}$). Changing now the
variables from $(a,\phi)$ to $(r,\theta)$ via the relations:
\begin{equation}
r=\sqrt{\frac{8}{3}}a^{3/2}\;\;\;\;\;\theta=\sqrt{\frac{3k\epsilon}{8}}\phi\;,
\label{tran1A}%
\end{equation}
the 2 dimensional metric (\ref{SF.52}) is given by
\begin{equation}
d{\hat{s}}^{2}=-dr^{2}+r^{2}d\theta^{2} \label{SF.56}%
\end{equation}
that is, $(r,\theta)$ are hyperbolic spherical coordinates in the 2
dimensional Minkowski space $\{a,\phi\}$. Next we introduce the new
coordinates $(x,y)$ with the transformation:%
\begin{align}
x  &  =r\cosh\;(\theta)\nonumber\label{trans}\\
y  &  =r\sinh\;(\theta)
\end{align}
which implies that eq.(\ref{SF.56}) becomes $d\hat{s}^{2}=-dx^{2}+dy^{2}$. We
also point here that
\begin{equation}
r^{2}=x^{2}-y^{2}\;\;\;\;\;\theta=\mathrm{arctanh}(y/x)\;. \label{tran1}%
\end{equation}
The scale factor ($a(t)>0$) is now given by:
\begin{equation}
a=\left[  \frac{3(x^{2}-y^{2})}{8}\right]  ^{1/3} \label{alcon}%
\end{equation}
which means that the new variables have to satisfy the following inequality:
$x\geq|y|$.

In the new coordinate system $(x,y)$ the Lagrangian (\ref{SF.50}) and the
Hamiltonian (\ref{SF.60e}) are written:
\begin{equation}
L=\frac{1}{2}\left(  \dot{y}^{2}-\dot{x}^{2}\right)  -V_{eff}(x,y)
\label{SF.60}%
\end{equation}
\begin{equation}
E=\frac{1}{2}\left(  \dot{y}^{2}-\dot{x}^{2}\right)  +V_{eff}(x,y)
\label{SF.60a}%
\end{equation}
where
\begin{equation}
V_{eff}(x,y)=\left(  x^{2}-y^{2}\right)  \left[  \tilde{V}\left(  \frac{y}%
{x}\right)  -\tilde{P}_{m}\right]  -3{\tilde K}_{3}\left(  x^{2}-y^{2}\right)
^{\frac{1}{3}} \;. \label{SF.60aa}%
\end{equation}
Note that we have used
\begin{equation}
\tilde{K}_{3}=K_{3}\left(  \frac{3}{8}\right)  ^{\frac{1}{3}}\;\;\;\;
\tilde{P}_{m}= \frac{3k}{8}P_{m}%
\end{equation}
and
\begin{equation}
\tilde{V}\left(  \theta\right)  = \frac{3k}{8}V(\theta)\;. \label{modpot}%
\end{equation}
If the matter pressure $P_{m}$ is constant then the Lagrangian (or
Hamiltonian) is time independent, thus the system is autonomous implying that
theorems 1 and 2 of \cite{Tsam10} apply. In this case we also have (for more
details see appendix A) that $\rho_{m}=\frac{|E|}{ka^{3}}-P_{m}$ which obeys
eq.(\ref{frie4}).

We now proceed in an attempt to provide the Lie point and Noether symmetries
of the current dynamical problem. Note, that for simplicity in the analytical
treatment below, unless explicitly stated, we consider spatially flat FRLW
($K_{3}=0$) scalar field models that include non-relativistic matter
($P_{m}=0$ with $\rho_{m}=\frac{|E|}{ka^{3}}$). Taking the latter into account
the effective potential of eq.(\ref{SF.60aa}) is given by
\begin{equation}
V_{eff}(x,y)=(x^{2}-y^{2})\tilde{V}(\frac{y}{x})=r^{2}\tilde{V}(\theta) \;.
\label{effective}%
\end{equation}
For this effective potential and with the aid of \cite{Leach} and
\cite{Tsam10} we find the following cases that admit Lie point and Noether symmetries:

\begin{itemize}
\item \textbf{Trivial Noether symmetries:} First of all, it is well known (see
for example \cite{Tsam10} and references therein) that for a general effective
potential $V_{eff}(x,y)$ we have only the Lie symmetry $\partial_{t}$.
Additionally, for effective potentials given by eq.(\ref{effective}) we have
that:
\[
V_{eff}(x,y)=x^{2}(1-\frac{y^{2}}{x^{2}})\tilde{V}\left(  \frac{y}{x}\right)
\]
hence the case of table 8, line 5 (with $d=0$) of \cite{Tsam10} applies and we
have the additional Lie symmetry $x\partial_{x}+y\partial_{y}$. A linear
combination between $\partial_{t}$ and $x\partial_{x}+y\partial_{y}$ provides
the following Lie symmetry:
\begin{equation}
X_{L}=c_{1}\partial_{t}+c_{2}\left(  x\partial_{x}+y\partial_{y}\right)  \;.
\label{SF.62}%
\end{equation}
From the corresponding tables of theorem 2 of \cite{Tsam10}, we see that in
this case there is only the trivial Noether symmetry\footnote{The operator
$\partial_{z}$ denotes $\partial/\partial z$. Also, there is a difference of
the results presented in \cite{Tsam10} due to the Lorentzian character of the
metric. This affects only the rotational part of the metric (non gradient
Killing vectors) and gives $y\partial_{x}+x\partial_{y}$ instead of the
Euclidean $y\partial_{x}-x\partial_{y}.$} $\partial_{t}$, whose Noether
integral is the Hamiltonian $E=$constant ($\partial_{t}E=0$). As it is
expected, this result implies that all the scalar field potentials described
in section 2.1 admit the trivial Noether symmetry, namely energy conservation,
as they should.

\item \textbf{Extra Noether symmetries:} Now we are looking for first
integrals beyond the standard one. In particular, we are interested to check
whether the scalar field potentials mentioned in this paper (see section 2.1)
can admit non-trivial Lie point/Noether symmetries. The agrument is simple: if
for a given potential we find extra Noether symmetries which are related with
conserved quantities then this particular model has an enhanced physical
meaning (see also \cite{Cap07}, \cite{Capa07},\cite{Cap08},\cite{Cap09}). From
the mathematical point of view the existence of extra integrals of motion
points the existence of an analytical solution (see next section). The novelty
in the current work is that we find that the exponential potential (see
eq.\ref{pot2}) and the UDM potential (see eq.\ref{unpot2}) can be clearly
distinguished from the other dark energy potentials because these are the only
potentials from the list presented in section 2.1, that accommodate extra
Noether symmetries.

\textbf{Hyperbolic - UDM Potential:} Generically, we use the following
potential:
\begin{equation}
\tilde{V}(\theta)=\frac{\omega_{1}\cosh^{2}\left(  \theta\right)  -\omega
_{2}\sinh^{2}\left(  \theta\right)  }{2} \label{hype1}%
\end{equation}
or
\begin{equation}
V_{eff}(x,y)=r^{2}\tilde{V}(\theta)=\frac{\omega_{1}x^{2}-\omega_{2}y^{2}}{2}
\label{hype2}%
\end{equation}
where we have used eqs.(\ref{trans}) and (\ref{effective}). The corresponding
Noether symmetries, $X_{n}$, are known (see for example \cite{Leach}). These
are (for $\omega_{1}\omega_{2}\neq0$ otherwise see appendix B):
\begin{align*}
X_{n_{1}}  &  =\partial_{t}~,~X_{n_{2}}=\sinh\left(  \sqrt{\omega_{1}%
}t\right)  \partial_{x}~,~X_{n_{3}}=\cosh\left(  \sqrt{\omega_{1}}t\right)
\partial_{x}\\
X_{n_{4}}  &  =\sinh\left(  \sqrt{\omega_{2}}t\right)  \partial_{y}%
~,~X_{n_{5}}=\cosh\left(  \sqrt{\omega_{2}}t\right)  \partial_{y}%
\end{align*}
The Noether integrals are the Hamiltonian and the quantities:%
\begin{align*}
I_{n_{2}}  &  =\sinh\left(  \sqrt{\omega_{1}}t\right)  \dot{x}-\sqrt
{\omega_{1}}\cosh\left(  \sqrt{\omega_{1}}t\right)  x\\
I_{n_{3}}  &  =\cosh\left(  \sqrt{\omega_{1}}t\right)  \dot{x}-\sqrt
{\omega_{1}}\sinh\left(  \sqrt{\omega_{1}}t\right)  x\\
I_{n_{4}}  &  =\sinh\left(  \sqrt{\omega_{2}}t\right)  \dot{y}-\sqrt
{\omega_{2}}\cosh\left(  \sqrt{\omega_{2}}t\right)  y\\
I_{n_{5}}  &  =\cosh\left(  \sqrt{\omega_{2}}t\right)  \dot{y}-\sqrt
{\omega_{2}}\sinh\left(  \sqrt{\omega_{2}}t\right)  y
\end{align*}
Obviously the UDM potential is a particular case of the current general
hyperbolic potential. Indeed for $\omega_{1}=2\omega_{2}$ and with the aid of
eqs.(\ref{tran1A}), (\ref{modpot}) we fully recover the UDM potential (see
section 2.1)
\begin{equation}
V(\phi)=V_{0}\left[  1+\cosh^{2}\left(  \frac{3k\epsilon}{8}\phi\right)
\right]  \label{pott}%
\end{equation}
where $V_{0}=\frac{4\omega_{2}}{3k}$ modulus a constant.

\textbf{Exponential Potential:} Here we provide for a first time (to our
knowledge) the Lie point and the Noether symmetries of the exponential
potential. Indeed from table 8, line 3 (with $d\neq0$) of \cite{Tsam10} one
can immediately see that
\[
V_{eff}(r,\theta)=r^{2}\tilde{V}(\theta)=r^{2}e^{-d\theta}\;.
\]
The corresponding Lie symmetry vector is:
\begin{equation}
X_{L}=2t\partial_{t}+\frac{4}{d}\left(  y\partial_{x}+x\partial_{y}\right)
\label{SF.63}%
\end{equation}
Concerning the Noether symmetries from table 14 of \cite{Tsam10} we find that
the Noether symmetry of the system for the potential $\tilde{V}(\theta
)=e^{-d\theta}$ is:
\begin{equation}
X_{n}=2t\partial_{t}+\left(  x+\frac{4}{d}y\right)  \partial_{x}+\left(
y+\frac{4}{d}x\right)  \partial_{y}\;. \label{SF.64}%
\end{equation}
In general the Noether integral for the vector $X_{n}=2t\partial_{t}+\eta
^{i}\partial_{i}$ is (see eq.66 of \cite{Tsam10}):%
\begin{equation}
I=2tE-\eta^{i}\hat{g}_{ij}\dot{x}^{i} \label{SF.67}%
\end{equation}
where $\eta^{i}=\left(  x+\frac{4}{d}y\right)  \partial_{x}+\left(  y+\frac
{4}{d}x\right)  \partial_{y}$. After some algebra we compute:%

\begin{align*}
\eta^{i}\hat{g}_{ij}\eta^{i}\dot{x}^{i}  &  =\left(
\begin{tabular}
[c]{ll}%
$x+\frac{4}{d}y~$ & $~~y+\frac{4}{d}x$%
\end{tabular}
\ \ \right)
\begin{pmatrix}
-1 & 0\\
0 & 1
\end{pmatrix}%
\begin{pmatrix}
\dot{x}\\
\dot{y}%
\end{pmatrix}
\\
&  =%
\begin{pmatrix}
-(x+\frac{4}{d}y)~ & ~\left(  y+\frac{4}{d}x\right)
\end{pmatrix}%
\begin{pmatrix}
\dot{x}\\
\dot{y}%
\end{pmatrix}
\\
&  =-\left(  x+\frac{4}{d}y\right)  \dot{x}+\left(  y+\frac{4}{d}x\right)
\dot{y}.
\end{align*}
i.e.
\begin{equation}
I=2tE+\left(  x+\frac{4}{d}y\right)  \dot{x}-\left(  y+\frac{4}{d}x\right)
\dot{y} \label{SF.68}%
\end{equation}
where $E$ is the Hamiltonian. Using ${\tilde{V}}=e^{-d\theta}$ together with
eq.(\ref{tran1A}) and eq(\ref{modpot}) we write the potential to its nominal
form (see section 2.1). This is
\begin{equation}
V(\phi)=V_{0}\mathrm{exp}\left(  -d\;\sqrt{\frac{3k\epsilon}{8}}\phi\right)
\label{pott}%
\end{equation}
where $V_{0}=\frac{8}{3k}$ modulus a constant\footnote{In the special case of
$d=2$, the system admits an additional Lie symmetry $\partial_{x}+\partial
_{y}$, with Noether integral $I=\dot{x}-\dot{y}.$}.
\end{itemize}

From now on, we focus on the exponential and the UDM potentials because they
contain non trivial integrals of motion, implying the existence of exact
analytical solutions (see next section). In section 4 we provide for a first
time (to our knowledge) such analytical solutions in the light of either
quintessence or phantom scalar field cosmologies that include also a
non-relativistic matter (cold dark matter) component.

\section{Analytical solutions in the flat scalar field cosmology}

In this section, we proceed in an attempt to analytically solve the
differential eqs.(\ref{klein1}) and (\ref{klein2}). We remind the reader that
this is possible because the dynamical system that includes either an
exponential or a UDM potential is autonomous and it admits extra Noether
symmetries (see previous section). Our aim is to derive the predicted time
dependence of the main cosmological functions namely $a(t)$ and $\phi(t)$ [and
thus of $H(t)$ and $w_{\phi}(t)$] in the scalar field cosmology.

\subsection{Analytical solutions of the hyperbolic potential}

Inserting eq.(\ref{hype2}) into eqs.(\ref{SF.50}),(\ref{SF.60}) the Lagrangian
and the Hamiltonian respectively become
\begin{equation}
L=\frac{1}{2}\left(  \dot{y}^{2}-\dot{x}^{2}\right)  -\frac{1}{2}\left(
\omega_{1}x^{2}-\omega_{2}y^{2}\right)  \label{Lagx}%
\end{equation}
\begin{equation}
E=\frac{1}{2}\left(  \dot{y}^{2}-\dot{x}^{2}\right)  +\frac{1}{2}\left(
\omega_{1}x^{2}-\omega_{2}y^{2}\right)  \;. \label{Hamx}%
\end{equation}
Technically speaking, in the new coordinate system our dynamical problem is
described by two independent hyperbolic oscillators and thus the system is
fully integrable. In particular, utilizing the Euler-Lagrange equations in the
new coordinate system the corresponding equations of motion can be written as
\begin{align*}
\ddot{x}-\omega_{1}x  &  =0\\
\ddot{y}-\omega_{2}y  &  =0
\end{align*}
where $\sqrt{\omega_{1}}$ and $\sqrt{\omega_{2}}$ are the oscillators'
''frequencies'' with units of inverse of time. It is routine to perform the
integration to find the analytical solutions:
\begin{align*}
x\left(  t\right)   &  =\sinh\left(  \sqrt{\omega_{1}}t+\theta_{1}\right) \\
y\left(  t\right)   &  =\sqrt{\frac{2E+\omega_{1}}{\omega_{2}}}\sinh\left(
\sqrt{\omega_{2}}t+\theta_{2}\right)
\end{align*}
\begin{align*}
a^{3}\left(  t\right)   &  =\frac{3}{8}[x^{2}\left(  t\right)  -y^{2}\left(
t\right)  ]\\
&  =\frac{3}{8}[\sinh^{2}\left(  \sqrt{\omega_{1}}t+\theta_{1}\right)
-\frac{2E+\omega_{1}}{\omega_{2}}\sinh^{2}\left(  \sqrt{\omega_{2}}%
t+\theta_{2}\right)  ]
\end{align*}
and
\begin{align*}
\phi\left(  t\right)   &  =\sqrt{\frac{8}{3k\epsilon}}\mathrm{arctanh}\left(
\frac{y}{x}\right) \\
&  =\sqrt{\frac{8}{3k\epsilon}}\mathrm{arctanh}\left[  \sqrt{\frac
{2E+\omega_{1}}{\omega_{2}}}\frac{\sinh\left(  \sqrt{\omega_{2}}t+\theta
_{2}\right)  }{\sinh\left(  \sqrt{\omega_{1}}t+\theta_{1}\right)  }\right]
\end{align*}
where $\theta_{1}$ and $\theta_{2}$ are the integration constants of the
problem. The constant $\theta_{1}$ is related to $\theta_{2}$ because at the
singularity ($t=0$), the scale factor has to be exactly zero. After some
algebra, we find that
\begin{align*}
\theta_{1}=\mathrm{ln}\left(  -\sqrt{\frac{2E+\omega_{1}}{\omega_{2}}}%
\sinh\theta_{2}+ \sqrt{\frac{2E+\omega_{1}}{\omega_{2}}\sinh^{2}\theta_{2}+1}
\right)  \;.
\end{align*}

We would like to remind the reader that the UDM dark energy model is recovered
for $\omega_{1}=2\omega_{2}$. Obviously, we can easily prove that the
concordance $\Lambda$-cosmology is a particular solution\footnote{In the
$\Lambda$CDM cosmology the scale factor is $a(t)=a_{0}\sinh^{2/3}%
(\omega_{\Lambda} t)$ where $\omega_{\Lambda}=3H_{0}\sqrt{\Omega_{\Lambda}}%
/2$, $a_{0}=(\Omega_{m}/\Omega_{\Lambda})^{1/3}$ and $\Omega_{\Lambda
}=1-\Omega_{m}$.} of the current hyperbolic potential with $\epsilon=1$,
$\omega_{1}=\omega_{2}=\omega_{\Lambda}$ and $(\theta_{1},\theta_{2})=(0,0)$.

Finally it is interesting to mention that the UDM cosmological model has been
tested against the latest cosmological data (SNIa and BAO) in Basilakos \&
Lukes \cite{BasLuk08}. In this paper the authors discussed the evolution of
matter perturbations as well as the spherical collapse model. They also
compared the UDM scenario with the traditional $\Lambda$-cosmology and they
found that the UDM scalar field model provides an overall (global and local)
dynamics which is in a fair agreement with that of the $\Lambda$-cosmology
although there are some differences especially at high redshifts.

\subsection{Analytical solutions of the exponential potential}

Now based on the exponential potential of eq.(\ref{pott}), we extend the
analytical solutions found by Russo \cite{Russo04} by taking into account the
presence of non-relativistic matter (cold dark matter), $\rho_{m}=\frac
{|E|}{ka^{3}}\ne0$. It is important to emphasize that Russo \cite{Russo04}
provided analytical solutions only in the context of quintessence DE
($\epsilon=1$) with $\rho_{m}=0$ while we solve analytically, for a first time
(to our knowledge), the current dynamical problem by treating dark energy
simultaneously either as quintessence or phantom with $\rho_{m}\ne0$. Since
for the current potential the $(x,y)$ coordinate system does not lead to an
analytical solution we are using the same methodology with that provided by
Russo \cite{Russo04}, we change variables from $(a,\phi)$ to $(u,v)$ according
to the transformations
\begin{align}
u  &  =\sqrt{\frac{3k\epsilon}{8}}\phi+\frac{1}{2}\ln\left(  a^{3}\right)
~\label{SFR.5}\\
v  &  =-\sqrt{\frac{3k\epsilon}{8}}\phi+\frac{1}{2}\ln\left(  a^{3}\right)
\;. ~ \label{SFR.6}%
\end{align}
Note that we have interchanged $(u,v)$ with respect to those of \cite{Russo04}%
. Inverting the above equations and using eq.(\ref{pott}) we get
\begin{equation}
a=e^{\frac{u+v}{3}}\;, \;\;\;\;\;\phi=\frac{1}{2}\sqrt{\frac{8}{3k\epsilon}%
}(u-v) \label{aaf}%
\end{equation}
and
\begin{equation}
V(u,v)=V_{0}e^{-d(u-v)/2} \;.
\end{equation}

In the new variables $(u,v)$ our Lagrangian (\ref{SF.50}) is written as
follows:
\begin{equation}
L=-e^{\left(  u+v\right)  }\left[  \frac{4}{3}\dot{u}\dot{v}+kV_{0}%
e^{-2K\left(  u-v\right)  }\right]  \;,~~K=\frac{d}{4} \label{SFR.7}%
\end{equation}
where $kV_{0}=8/3$. The next step is to consider a change in the time
coordinate as follows:
\begin{equation}
\frac{d\tau}{dt}=\sqrt{\frac{3kV_{0}}{4}}e^{-K\left(  u-v\right)  }
\label{SFR.8}%
\end{equation}
which implies:
\begin{align*}
\dot{v}  &  =\frac{dv}{dt}=\frac{dv}{d\tau}\frac{d\tau}{dt}=v^{\prime}%
\sqrt{\frac{3kV_{0}}{4}}e^{-K\left(  u-v\right)  }\\
\dot{u}  &  =\frac{du}{dt}=\frac{du}{d\tau}\frac{d\tau}{dt}=u^{\prime}%
\sqrt{\frac{3kV_{0}}{4}}e^{-K\left(  u-v\right)  }%
\end{align*}
where $u^{\prime}=\frac{du}{d\tau}$ and $v^{\prime}=\frac{dv}{d\tau}$.
Obviously using the latter transformations, eq.(\ref{SFR.7}) and
eq.(\ref{SFR.8}) the action given by eq.(\ref{action}) takes the form:%
\begin{equation}
S=-\sqrt{\frac{4kV_{0}}{3}}\int d^{3}x\;d\tau e^{\left(  u+v\right)
}e^{-K\left(  u-v\right)  }\left(  u^{\prime}v^{\prime}+1\right)  .
\label{SFR.9}%
\end{equation}
Now varying the action we arrive at
\begin{align}
u^{\prime\prime}+\left(  1-K\right)  u^{\prime2}-\left(  1+K\right)   &
=0\label{SFR.11}\\
v^{\prime\prime}+\left(  1+K\right)  v^{\prime2}-\left(  1-K\right)   &  =0.
\label{SFR.12}%
\end{align}
Obviously, in the latter equations the variables $u,v$ decouple. In these
variables, the Hamiltonian
of the system becomes:
\begin{equation}
E=e^{\left(  u+v\right)  } e^{-K\left(  u-v\right)  } (u^{\prime}v^{\prime}-1)
\label{SFR9.a}%
\end{equation}
where $E \neq0$ (or $\rho_{m} \ne0$). As we have already stated the above
system of equations has been derived also by \cite{Russo04} in the case of
quintessence dark energy ($\epsilon=+1)$, and it is solved only for $E=0$ (see
appendix A). Here we prove that the same equations are valid also in the case
of phantom dark energy in which the scalar field is imaginary, however the
potential, the scale factor and the FRLW metric are real as they should.

We conclude that the FRLW metric (using eqs. \ref{SFR.8} and \ref{aaf}) in the
coordinates $(\tau,x^{\mu})$ is:
\begin{equation}
ds^{2}=-\frac{4}{3kV_{0}}\;e^{4K\sqrt{\frac{3k\epsilon}{8}}\phi(\tau)}%
d\tau^{2}+a^{2}(\tau)dx^{i}dx_{i}\;. \label{metric1}%
\end{equation}
Bellow we provide analytical solutions for the two different cases.

\subsubsection{Case $K=1$}

In this case the system of eqs.(\ref{SFR.11}), (\ref{SFR.12}) and
(\ref{SFR9.a}) becomes:%
\begin{align*}
u^{\prime\prime}-2  &  =0\\
v^{\prime\prime}+2v^{\prime2}  &  =0
\end{align*}%
\[
e^{2v}\left(  u^{\prime}v^{\prime}-1\right)  =E
\]
and the solution is:%
\begin{align*}
u\left(  \tau\right)   &  =\tau^{2}+c_{1}\tau\\
v\left(  \tau\right)   &  =\frac{1}{2}\ln\left(  2c_{3}\tau\right)
\end{align*}
where the constants are related by the constraint:%
\[
E=c_{3}c_{1}\rightarrow c_{3}=\frac{E}{c_{1}}.
\]
Replacing we find:%
\begin{align*}
u\left(  \tau\right)   &  =\tau^{2}+c_{1}\tau\\
v\left(  \tau\right)   &  =\frac{1}{2}\ln\left(  \frac{2E}{c_{1}}\tau\right)
\;.
\end{align*}
We note that the solution depends on one arbitrary parameter $c_{1}\neq0.$ If
we choose
$c_{1}=E$ then we have the solution:%
\begin{align*}
u\left(  \tau\right)   &  =\tau^{2}+E\tau\\
v\left(  \tau\right)   &  =\frac{1}{2}\ln\left(  2\tau\right)
\end{align*}
from these follows:%
\begin{align*}
a^{3}\left(  \tau\right)  =\sqrt{2\tau} e^{\tau^{2}+E\tau}%
\end{align*}
\begin{align*}
\phi(\tau)=\frac{1}{4}\sqrt{\frac{8}{3k\epsilon}} \left[  2\tau^{2}+2E\tau
-\ln\left(  2\tau\right)  \right]  \;.
\end{align*}

\subsubsection{Case $~K\ne1$}

In this case the solution of the system is:
\begin{align*}
u\left(  \tau\right)   &  =-\frac{1}{2\left(  K-1\right)  }\ln\left[
Cc_{3}^{2}\;\mathrm{Sinn}^{2}\left(  \sqrt{|K^{2}-1|}\tau+\theta_{1}\right)
\right] \\
v\left(  \tau\right)   &  =\frac{1}{2\left(  1+K\right)  }\ln\left[
C^{-1}c_{1}^{2}\;\mathrm{Sinn}^{2}\left(  \sqrt{|K^{2}-1|}\tau+\theta
_{1}\right)  \right]
\end{align*}
where $C=\frac{|K-1|}{1+K}$,
\begin{equation}
\mathrm{Sinn}\omega=\left\{
\begin{array}
[c]{cc}%
\mathrm{sin}\omega & \mbox{$K>1$}\\
\mathrm{sinh}\omega & \mbox{$\;\;\;0<K<1$}
\end{array}
\right.  \label{SSINN}%
\end{equation}
and $\theta_{1}$ being the phase constant. Without loosing the generality we
can select $\theta_{1}=0$. Also, the two constants of integration satisfy the
condition: $E=c_{3}c_{1}$. Next we may choose $c_{3}=E\neq0$ and have the
solution:
\begin{align*}
u\left(  \tau\right)   &  =-\frac{1}{2\left(  K-1\right)  }\ln\left[
CE^{2}\;\mathrm{Sinn}^{2}\left(  \sqrt{|K^{2}-1|}\tau\right)  \right] \\
v\left(  \tau\right)   &  =\frac{1}{2\left(  1+K\right)  }\ln\left[
C^{-1}\;\mathrm{Sinn}^{2}\left(  \sqrt{|K^{2}-1|}\tau\right)  \right]  \;.
\end{align*}
Finally, we transform this solution to the coordinates $a\left(  \tau\right)
,\phi\left(  \tau\right)  $. Doing so we obtain
\[
a^{3}\left(  \tau\right)  =C^{-\frac{K}{K^{2}-1}}|E|^{\frac{1}{K-1}%
}\mathrm{Sinn}\left(  \sqrt{|K^{2}-1|}\tau\right)  ^{-\frac{2}{K^{2}-1}}
\]%
\[
\phi\left(  \tau\right)  =-\frac{1}{2}\sqrt{\frac{8}{3k\epsilon}}\ln\left[
C^{-\frac{1}{K^{2}-1}}|E|^{\frac{1}{K-1}}\mathrm{Sinn}\left(  \sqrt{|K^{2}%
-1|}\tau\right)  ^{\frac{2K}{K^{2}-1}}\right]  \;.
\]

It is interesting to mention that in the case of $0<K<1$ at late enough times
we obtain that $\tau\sim\mathrm{ln}t$ and thus the scale factor evolves as
$a(t)\propto t^{2/3\sqrt{1-K^{2}}}$. The current solution of the scale factor
can provide, a recent cosmic acceleration ($\ddot{a}(t)>0$) for $K\in
(\frac{\sqrt{5}}{3},1)$.

\section{Conclusions}

In this paper we propose to use a theoretical model-independent criterion,
based on first integrals of motion, usually named Noether symmetries in order
to discriminate the dark energy (quintessence or phantom) models within the
context of scalar field FLRW cosmology. This is possible via the geometrical
symmetries of the space-time in which both gravity and dark energy live. In
particular, following the general methodology of \cite{Tsam10} (see also the
references therein), the Noether symmetries are computed for 9 distinct
accelerating cosmological scenarios that contain a homogeneous scalar field
associated with different types of potentials. Note that the free parameters
of the dark energy models studied here can be constrained by using the current
cosmological data. In particular, one has to perform a joint likelihood
analysis utilizing for example the SNIa data \cite{Hic09}, the shift parameter
of the Cosmic Microwave Background (CMB)\,\cite{komatsu08} and the observed
Baryonic Acoustic Oscillations (BAOs; \cite{Eis05}). Such an analysis is in
progress and will be published elsewhere.

The main results of the current paper can be summarized in the following
statements (see sections 3.2 and 4):

\begin{itemize}
\item We verify that all the scalar field potentials, studied here, admit the
trivial first integral, namely energy conservation as they should.

\item We find that the exponential and the unified dark matter potentials
occupy an eminent position in the scalar field potentials hierarchy, being the
potentials that admit extra integrals of motion, and therefore appear to be
promising candidates for describing the physical properties of dark energy as
well as extracting useful cosmological information. The existence of the new
Noether integrals can be used to simplify the system of differential equations
(equations of motion) as well as to determine the integrability of the system.

\item Based on the exponential and hyperbolic potentials we find that the main
cosmological functions, such as the scale factor of the universe, the scalar
field, the Hubble expansion rate and the metric of the FRLW space-time are
provided analytically.
\end{itemize}

In a future work we plan to apply the same approach also to $f(R)$
cosmological models.

\vspace{0.5cm} \textbf{Acknowledgments.} We would like to thank M. Plionis and
L. Perivolaropoulos for useful comments and suggestions.

\appendix

\section{Matter density versus Hamiltonian}

We remind the reader that if the matter pressure is constant then the
dynamical system described by the general Lagrangian of eq.(\ref{SF.50}) is
autonomous. Therefore, one can easily prove that $\rho_{m}=\frac{|E|}{ka^{3}%
}-P_{m}$. Indeed, utilizing eq.(\ref{frie1}) and eq.(\ref{SF.60e}) we have
after some simple algebra that:
\begin{align*}
\left(  \frac{\dot{a}}{a}\right)  ^{2}+\frac{K_{3}}{a^{2}}  &  =\frac{k}%
{3}\left(  \rho_{m}+\frac{1}{2}\epsilon\dot{\phi}^{2}+V(\phi)\right)
\Rightarrow
\end{align*}
\begin{align*}
a\dot{a}^{2}+K_{3} a  &  =\frac{k}{6}a^{3}\epsilon\dot{\phi}^{2}+a^{3}\frac
{k}{3}\rho_{m}+\frac{k}{3}a^{3}V\left(  \phi\right)  \Rightarrow\\
\end{align*}
\begin{align*}
-a\dot{a}^{2}+\frac{k}{6}a^{3}\epsilon\dot{\phi}^{2}+ \frac{k}{3}a^{3}\left[
V\left(  \phi\right)  -P_{m}\right]  -K_{3} a=-\frac{k}{3}a^{3}\left[
P_{m}+\rho_{m}\right] \\
\end{align*}
or
\begin{align*}
E  &  =-k a^{3}\left[  P_{m}+\rho_{m}\right]
\end{align*}
and thus:
\[
\rho_{m}=\frac{|E|}{ka^{3}}-P_{m}
\]
which satisfies eq.(\ref{frie4}). Note that the inequality $\rho_{m}\ge0$
points that $E\le0$. The case of non-relativistic matter $P_{m}=0$ implies
$\rho_{m}(a)=\frac{|E|}{ka^{3}}$.

\section{Solutions of Harmonic Oscillator coupling with a free particle}

Here we consider either $\omega_{1}\ne0$, $\omega_{2}=0$ or $\omega_{1}=0$,
$\omega_{2}\ne0$. For the latter case one may checks \cite{Cap09}. In both
cases the system is equivalent to a pair of two dynamical systems a simple
harmonic oscillator and a free particle. The Lie point and the Noether
symmetries of the current system can be found in \cite{Blum}. As an example
for $\omega_{1}\ne0$, $\omega_{2}=0$ equations (\ref{Lagx}) and (\ref{Hamx})
become
\[
L=\frac{1}{2}\left(  \dot{y}^{2}-\dot{x}^{2}\right)  -\frac{\omega_{1}x^{2}%
}{2}
\]%
\[
E=\frac{1}{2}\left(  \dot{y}^{2}-\dot{x}^{2}\right)  + \frac{\omega_{1}x^{2}%
}{2}
\]
and
\[
\ddot{x}-\omega_{1}x=0 \;\;\;\; \ddot{y}=0 \;.
\]
The solution of the above system is
\begin{align*}
x\left(  t\right)   &  =\text{\textrm{sinh}}\left(  \sqrt{\omega_{1}}%
t+{\theta}_{1}\right) \\
y\left(  t\right)   &  =\sqrt{2E+\omega_{1}}t
\end{align*}
or
\begin{align*}
a^{3}\left(  t\right)   &  = \frac{3}{8}[x^{2}\left(  t\right)  -y^{2}\left(
t\right)  ]\\
&  =\frac{3}{8}[\text{\textrm{sinh}}^{2}\left(  \sqrt{\omega_{1}}t+ \theta
_{1}\right)  -\left(  2E+\omega_{1}\right)  t^{2}]\\
\phi\left(  t\right)   &  =\sqrt{\frac{8}{3k\epsilon}}\mathrm{arctanh}\left(
\frac{t\sqrt{2E+\omega_{1}}}{\text{\textrm{sinh}}\left(  \sqrt{ \omega_{1}%
}t+\theta_{1}\right)  }\right)  \;.
\end{align*}
Note that due to $a(0)=0$ we have $\theta_{1}=0$.

\section{Exponential Potential versus empty space $\rho_{m}=0$ (or $E=0$)}

In this appendix we would like to give the reader the opportunity to
appreciate the fact that our solutions provided in section 4 can be viewed as
an extension of those found by Russo \cite{Russo04} for the quintessence
($\epsilon=1$) dark energy with $\rho_{m}=0$ (or $E=0$). For either
quintessence or phantom dark energy, the system of equations (\ref{SFR.11}),
(\ref{SFR.12}) and (\ref{SFR9.a}) which we have to solve is:
\begin{align*}
u^{\prime\prime}+\left(  1-K\right)  u^{\prime2}-\left(  1+K\right)   &  =0\\
v^{\prime\prime}+\left(  1+K\right)  v^{\prime2}-\left(  1-K\right)   &  =0
\end{align*}%
\begin{equation}
E=(u^{\prime}v^{\prime}-1)e^{\left(  u+v\right)  } e^{-K\left(  u-v\right)  }
\label{EENER}%
\end{equation}
Due to the fact that $E=0$ eq.(\ref{EENER}) takes the form $u^{\prime
}v^{\prime}=1$. Thus we consider the following cases:

\textbf{Case 1:} For $K=1$ the solution is:
\begin{align*}
u\left(  t\right)   &  =\tau^{2}%
\end{align*}
\begin{align*}
v\left(  t\right)   &  =\frac{1}{2}\ln\left(  2\tau\right)
\end{align*}
or
\begin{align*}
a^{3}\left(  \tau\right)  =\sqrt{2\tau} e^{\tau^{2}}%
\end{align*}
\begin{align*}
\phi\left(  \tau\right)  =\frac{1}{4}\sqrt{\frac{8}{3k\epsilon}} \left[
2\tau^{2}-\ln\left(  2\tau\right)  \right]  \;.
\end{align*}

\textbf{Case 2:} For $K\ne1$ the solution is:
\begin{align*}
u\left(  \tau\right)   &  =-\frac{1}{\left(  K-1\right)  }\ln\left[
\mathrm{Sinn}\left(  \sqrt{|K^{2}-1|}\tau+\theta_{1}\right)  \right] \\
v\left(  \tau\right)   &  =\frac{1}{\left(  K+1\right)  }\ln\left[
\mathrm{Coss}\left(  \sqrt{|K^{2}-1|}\tau+\theta_{1}\right)  \right]
\end{align*}
or (for $\theta_{1}=0$)
\begin{align*}
a^{3}\left(  \tau\right)  = \frac{\mathrm{Coss}\left(  \sqrt{K^{2}-1}%
\tau\right)  ^{\frac{1}{\left(  K+1\right)  }} } {\mathrm{Sinn}\left(
\sqrt{|K^{2}-1|}\tau\right)  ^{\frac{1}{\left(  K-1\right)  }} }%
\end{align*}
\begin{align*}
\phi\left(  \tau\right)  = -\frac{1}{2}\sqrt{\frac{8}{3k\epsilon}} \ln\left[
a^{3}(\tau) \mathrm{Sinn}\left(  \sqrt{|K^{2}-1|}\tau\right)  ^{\frac
{2}{\left(  K-1\right)  }} \right]
\end{align*}
where the quantity $\mathrm{Sinn}$ is given by (eq.\ref{SSINN}) and
\[
\mathrm{Coss}\omega=\left\{
\begin{array}
[c]{cc}%
\mathrm{cos}\omega & \mbox{$K>1$}\\
\mathrm{cosh}\omega & \mbox{$\;\;\;0<K<1$.}
\end{array}
\right.
\]
We point that for $\epsilon=1$ the current solutions coincide (modulus some
constants) those of \cite{Russo04}. In the case of phantom dark energy
($\epsilon=-1$) the scalar field is imaginary, however the potential, the
scale factor and the metric of the space-time are real as they should.


\end{document}